%% file: map.tex
\pdfoutput=1
\DocumentMetadata{}
\documentclass[acmsmall,manuscript,nonacm]{acmart}
\setcopyright{none}

\usepackage{booktabs}   
\usepackage{subcaption}
\usepackage{enumerate}
\usepackage{latexsym}
\usepackage{amsfonts}
\usepackage{amstext}
\usepackage{array}
\usepackage{amsmath}
\usepackage{amsthm}
\usepackage{bm}
\usepackage{hyperref}
\usepackage{float}
\usepackage{caption}
\usepackage{framed}
\usepackage{color}
\usepackage{mathtools}
\usepackage[normalem]{ulem}
\usepackage{titlesec}

\DeclareMathOperator*{\argmax}{arg\,max}

\newtheorem{thm}{Theorem}[section]
\newtheorem{defi}[thm]{Definition}

\newcommand{\stkout}[1]{\ifmmode\text{\sout{\ensuremath{#1}}}\else\sout{#1}\fi}

\newcommand{\cut}[2]{{\color{orange}#1}{\color{green}#2}}

\newcommand{\removeforfinal}[1]{#1}
\newcommand{\removeforshortfinal}[1]{#1}

\makeatletter
\renewcommand{\paragraph}{%
  \@startsection{paragraph}{4}%
  {\z@}{1.4ex \@plus 1ex \@minus .2ex}{-1em}%
  {\normalfont\normalsize\bfseries}%
}
\makeatother

\renewcommand{\cut}[2]{#1}
\renewcommand{\removeforfinal}[1]{}

\begin{document}

\title[On Heuristic Models, Assumptions, and Parameters]{On Heuristic Models, Assumptions, and Parameters} 

\author{Samuel Judson}
\authornote{Work completed while author was at Yale University.}
\orcid{0000-0003-1270-6601}
\email{sam@nexus.xyz}             
\affiliation{
  \institution{Nexus}            
  \country{USA}
}

\author{Joan Feigenbaum}
\orcid{0000-0002-3735-6022}
\email{joan.feigenbaum@yale.edu}    
\affiliation{
  \institution{Yale University} 
  \country{USA}
}

\renewcommand{\shortauthors}{Judson and Feigenbaum}

\begin{abstract}
Insightful interdisciplinary collaboration is essential to the principled governance of technology. When such efforts address the interaction between computation and society, they often focus on modeling, the process by which computer scientists formally define problems in order to enable algorithmic solutions. But modeling is a multifaceted and inherently imperfect process. Especially in interdisciplinary work, it often receives uneven scrutiny because of the practical challenges of communicating complex technical details to non-experts. We argue that there is an underappreciated if loose family of obscure and opaque technical caveats, choices, and qualifiers that the social effects of computing can depend just as much on as far more heavily scrutinized modeling choices. These artifacts are often used by researchers to paper over the incomplete theoretical foundations of computing or to burden shift responsibility for the impact of normative design decisions. Further, their nuanced technical nature often complicates thorough sociotechnical scrutiny of the discretionary decisions made to manage them. We describe three specific classes of such objects: heuristic models, assumptions, and parameters. We raise six reasons these objects may be hazardous to comprehensive analysis of computing and argue they deserve deliberate consideration as researchers explain scientific work.
\end{abstract}



\maketitle

\input{./content.tex}

\begin{acks}         

The work of the first author was supported in part by the National Defense Science and Engineering Graduate (NDSEG) Fellowship and by US National Science Foundation grant CCF-2131476. The work of the second author was supported in part by US National Science Foundation grant CCF-2131356. The first author completed substanative work on this paper while at Yale University, and its content is not associated with his current affiliation of Nexus. The second author holds concurrent appointments as the Grace Murray Hopper Professor of Computer Science at Yale University and as an Amazon Scholar. This paper describes work performed at Yale and is not associated with Amazon.

\end{acks}
\bibliography{./map}

\end{document}

%% file: content.tex
\section{Introduction}\label{sec:intro}

Computing is integral to modern society. Its complex, automated, and often autonomous nature, however, makes its responsible use a significant social challenge. In particular, the principled governance of computing suffers from a problem of collective action. Those with the deepest technical understanding of computation, such as scientists and engineers, are rarely those best equipped -- let alone authorized to -- govern its use, compared to policymakers, judges, lawyers, regulators, and a democratic society at large. Perhaps inevitably then, interdisciplinary efforts to govern technology have a tendency to work off of shared knowledge bases that emphasize informal, often analogical, reasoning over precise mathematical reasoning. Interdisciplinary research aimed at the governance of algorithmic markets is, for instance, almost certain to appeal to or assume an intuitive understanding of economic rationality, far more than it is likely to delve into a rigorous mathematical exposition invoking the theories of calculus, probability, and computational complexity. The hope is, in some sense, that an informal understanding guided by analogy can be enough -- and in many cases that belief may very well hold. But it will not always. In this paper, we argue there is an underappreciated family of often obscure and opaque formalisms -- \textit{heuristic models}, \textit{assumptions}, and \textit{parameters}, all in common use throughout computer science -- that form an outsized hazard to effective interdisciplinary collaboration intended to support the principled governance of computing.

Consider the case of \emph{differential privacy} (DP)~\cite{dwork2006calibrating, dwork2014algorithmic} as used by the United States Census Bureau for its 2020 decennial census. By statute, the Census Bureau is prevented from producing `any publication whereby the data furnished by any particular establishment or individual ... can be identified'~\cite{cornellLII}. Despite this restriction, the Census Bureau produces numerous reports that are widely used by legislators, regulators, policymakers, and researchers at all levels both within and without government~\cite{nyt, kenny2021impact}. Computer scientists have shown that robust anonymization of data is an
extraordinarily difficult problem~\cite{dinur2003revealing, narayanan2008robust, ohm2009broken}, with years of intensive research to improve upon heuristics culminating in differential privacy as the first theoretically sound approach. After a thorough review, the Census Bureau chose to adopt it for the 2020 census in order to provide the strongest possible assurance of compliance with their privacy mandate~\cite{
abowd2019economic, garfinkel2018issues, dpriv, dpriv2}. This decision was and remains contentious, spurring both litigation~\cite{avcom, cresp} and extensive debate~\cite{cb, cb2, nyt, int, ruggles2021role, santos2020differential,wood2017differential,hotz2022balancing,cohen2020towards,abowd2019economic, garfinkel2018issues, kenny2021impact}. That broader debate invoked many nuanced scientific and social principles -- individual privacy and the social compact,\cut{ the utility vs.~privacy tradeoff inherent in population statistics,}{} federalism, bureaucratic governance and administrative law, database reconstruction and calibrated noise -- with varying degrees of mathematical sophistication. But the defense of deploying differential privacy for disclosure avoidance ultimately rested in large part on an otherwise unassuming real number: $19.61$.

Differential privacy is a framework for calibrating a tradeoff between the utility of statistics and the privacy of their underlying data. At one extreme, it permits the choice of maximum utility with no privacy and, at the other, perfect privacy with no utility. In practice, selection of a real-valued parameter $\epsilon$ concretely fixes the privacy loss permitted, and with it the corresponding loss in accuracy, to somewhere between the two extremes. The Census Bureau set global $\epsilon = 19.61$~\cite{cb, cb2}. So while some dispute focused on the distinctive method by which DP protects privacy or on procedural delays its use may have caused, many critiques simply advanced the claim that the US Census Bureau had unreasonably accepted a significant loss of accuracy for, at best, an insufficient gain in privacy. The Census Bureau and differential privacy researchers put forth compelling responses to these criticisms. But it was here that the oft-invoked (\textit{e.g.},~\cite{wood2017differential, dwork2014algorithmic, dwork2019differential}) analogical explanation of differential privacy -- that it enables paying with privacy for utility out of a socially-determined budget -- reached its limit. In practice, the most vocal criticisms were levelled not at the idea of modeling statistical privacy in terms of adversarial reconstruction of an individual's data nor at the normative design decision to pay for social utility with individual privacy. Rather, the criticisms targeted the allegedly exorbitant cost accepted by the Census Bureau. And while the need to select an $\epsilon$ can quite naturally be conveyed to a non-technical audience~\cite{wood2017differential}, the impact of any particular choice cannot -- yet it was from the implications of fixing the specific choice of $\epsilon = 19.61$ that much of the resulting controversy flowed. For analysis of the practical impact of its use with the 2020 census, an intuitive understanding of differential privacy is not necessarily enough.

The example of the $\epsilon$ of differential privacy is far from unique within computer-scientific research. Computing is rife with small and unintuitive technical details that nonetheless have an outsized impact on the societal effects of a technology. And while effective governance is ultimately produced and managed by lawyers, regulators, policymakers, and other practitioners that are experts in governance, they most often lack esoteric mathematical and computer-scientific knowledge. This limitation often also extends as well to the legal, policy, and social-science researchers whose findings support effective policy- and rule-making. On the other hand, computer scientists and engineers themselves rarely possess the legal and policy sophistication to propose feasible and comprehensive sociotechnical mechanisms of governance. As a consequence, their work often falls into the trap of solutionism~\cite{selbst2019fairness}. The need for both technical and societal expertise has motivated both the deeper integration of computer scientists into the law-making and policy-making processes and the development of an immense body of interdisciplinary research reaching across almost every subfield of computer science: cryptography and privacy, artificial intelligence and machine learning, human-computing interaction, program verification and analysis, computer graphics, and more. This research has dual but complementary goals. At the least it aims to document and communicate the particular nature and peculiar effects of computing in order to effect change through governance by law and policy. When possible, it further aims to design bespoke algorithmic methods with an eye towards technical mitigations or `solutions' for societal problems -- whether those problems originate within or without computing itself.

In this paper, we argue that some of those peculiar technical details present often underappreciated hazards to both goals of work at the intersection of computing and society. We classify three technically distinct but philosophically similar objects used within computer science:
\begin{enumerate}[i]
    \item proofs that hold only over \emph{heuristic models};
    \item technical \emph{assumptions} of believed rather than proven truth; and
    \item numerical \emph{parameters} that encode complex tradeoffs through often deceptively simple choices.
\end{enumerate}
Cumulatively, we deem these \emph{HMAPs}. Each is the product of norms developed by computer scientists in order to represent complex and messy social and physical systems within the sterile mathematical language required of algorithm design and analysis. These often unassuming mathematical objects can greatly influence -- or even entirely define -- the real-world consequences of deploying a technology, while simultaneously being poorly understood by technical experts and obscure and opaque to practitioners who must govern their use in order to promote the social interest and welfare.

Further, we contend that HMAPs often receive insufficient attention in interdisciplinary discussion, which may undercut legal, regulatory, and policymaking processes governing computing. We therefore aim to motivate further recognition and discussion about HMAPs in law- and policy-oriented work produced by technical researchers. Through examples drawn from cryptography, privacy, and machine learning in particular, we argue such work may be essential to reducing friction within the collective action of technologists and practitioners working towards effective governance of computing, as well as to navigating the particular hazards to those efforts that HMAPs present.

\paragraph{The Nature of HMAPs.} As with any science, computer-science researchers and technical engineers working on `practical' problems must build a formal model of a social or physical process. This model is inherently lossy -- as the adage goes, `all models are wrong, but some are useful'~\cite{box1979robustness, smith1985limits}. But computer science has a particular predilection for relying on \emph{heuristic models} used not because the world is messy, but because the mathematics are. The compromise inherent in heuristic models can limit the applicability of algorithmic results based on them for reasons not immediately visible to anyone without expertise in the specific mathematical methods. Technical \emph{assumptions} allow computer scientists to assert and prove algorithmic behaviors that are believed true but not known to be. They are the shifting foundations of many computer-science results, and a finding of their incompleteness or invalidity can rapidly shift a field -- an acute threat that hangs continually over cryptography in particular, because the invalidity of a hardness assumption can undermine security arguments reliant upon it, in some cases leading to practical attacks. Numerical \emph{parameters} allow for the offloading of complex questions of social choice and technical efficacy onto opaque valuations that often receive less attention than the algorithmic methods whose effects rely on their careful choice.

The choice of $\epsilon$ for DP is far from unique in how arcane technical detail can drive social consequences. These details often require significant technical expertise to understand -- or even to discern the relevancy of in the first place. As another example, in cryptography it is careful problem modeling that motivates security guarantees; captures participants, adversarial intent, and computational capacity; and justifies the overall conclusion that a construction is secure. Any such conclusion, however, may be conditioned on the use of adequate key sizes and the assumed hardness of a computational problem~\cite{goldreich2009foundations, katz2014introduction}. Both of these qualifiers have been exploited to diminish the practical security of theoretically secure systems. Export-grade cryptography was weakened through legally mandated short key lengths~\cite{abelson2015keys, diffie2001export}, while (almost certainly) malicious parameter generation circumvented the security arguments justifying the DUAL\_EC\_DRBG pseudorandom-number generator~\cite{brown2007security, green2013many}. In machine learning, problem modeling legitimizes a data universe as adequately representing a social or physical system so that training and validating against it will be useful in deployment~\cite{nickel2024no}. It also justifies particular choices of hypothesis class and loss function, and the systemic interpretation of computational outputs. It shapes understanding and implementation of desirable qualities such as fairness or robustness, and it gives confidence that a resulting trained model will be accurate when deployed~\cite{mulligan2019thing, selbst2019fairness, shalev2014understanding}. However, particular definitions of fairness can have social impact dependent upon a choice of parameters intended to encode a desired equity, in a manner reminiscent of, or even directly descended from, the $\epsilon$ of differential privacy~\cite{celis2017ranking, dwork2012fairness}. In this work, we frequently proceed by example, primarily drawn from origins in cryptography and privacy or artificial intelligence and machine learning.

\paragraph{The Importance of HMAPs.} Debates over the societal and legal governance of technology almost inevitably focus on questions of modeling: how scientists and engineers represent social or physical problems in order to formulate technical solutions. But modeling in an involved and intricate process, and because the more fine-grained a model is the more complex it may be to use and reason about, there is a natural inclination among scientists to mitigate complexity. Computer scientists, in particular, rely upon a proven suite of techniques for reducing normative choices and tradeoffs down to parameters and for confining caveats and qualifiers -- often tied back to the incomplete complexity-theoretic foundations of the field -- to careful reliance upon imperfect models and assumptions. When computer scientists aim only towards traditional goals of algorithmic correctness and efficiency, these methods are mostly of just scientific interest. However, for much of the research community, correct and efficient is an insufficient standard, and computation must further be, \emph{e.g.}, accountable, fair, just, explainable, interpretable, moral, legal, or politically sensible~\cite{abebe2020roles, abelson2015keys, barocas2016big, chaum1985security, doshi2017towards, kroll2016accountable, lessig2009code, ohm2009broken, cryptoeprint:2015:1162, rudin2019stop, malik2020hierarchy, mulligan2019thing, nissenbaum1996accountability, scheffler2021protecting, selbst2019fairness, weitzner2008information}. In light of such broader norms, this otherwise innocuous suite of techniques presents a unique challenge for interdisciplinary research at the intersection of computer science with law, policy, and society, let alone for the actual practice of technological governance. As our examples demonstrate, parameters and assumptions enable computations to have consequences that arise not just from how a problem is modeled or the basic mechanism of a proposed solution, but rather from the obscure and opaque technical details of particular algorithmic techniques and their theory. For DP, the basic principle of trading some statistical accuracy to protect individual privacy can be approachable and intuitive for non-technical audiences~\cite{wood2017differential}. Yet, the implications to American society and governance of the particular tradeoff encoded by $\epsilon = 19.61$ have proven far more muddied and contentious.

The ongoing groundswell of interest -- both purely technical and interdisciplinary -- in the interaction between computation and society motivates a critical look at what form these techniques take, and the implications they can carry for the understanding and governance of computing by practitioners, especially those without deep technical expertise in computer science. Although technical expertise may not be essential to diagnose the ramifications of technology, it is required in order to locate their root technical causes or when proposing technically involved mechanisms for \emph{ex ante} mitigation of or \emph{ex post} accountability for computing harms. Any `solution' that cannot be well defined or is uncomputable, intractable, or inconsistent with the algorithmic methods used is simply a non-starter. The obscurity of HMAPs make them potentially more insidious hurdles. As noted, a knowledgeable and engaged policy debate on the \emph{cost} of DP for census anonymity is much more difficult than one on its \emph{use}, because the algorithmic technique of DP itself is much more approachable than the choice of $\epsilon$.

The barriers to understanding that HMAPs create are just one example of an essential and widely discussed difficulty with bridging gaps in knowledge and methods in interdisciplinary work between the sciences and other fields, perhaps most famously expounded on by C.P. Snow in his iconic \emph{Two Cultures} lecture and book~\cite{snow1959two} and continually revisited since by computer, physical, and social scientists, as well as law and policy scholars~\cite{bauer1990barriers,schuck1993multi,mulligan2019thing}. Not only the difficulty but also the urgency of effective communication of the technical underpinnings of computing harms is itself far from novel for the intersection of science and society. It often arises when foundational technologies `escape the laboratory' and begin to reshape society -- as occurred, \emph{e.g.}, when physicists greatly influenced public policy and morality in the atomic age~\cite{reman}.

More specifically, the nature and implications of HMAPs touch on common themes in the philosophy of science~\cite{sep-models-science}, to the extent that similar concerns have been raised about even some of the same objects within domains adjacent to computing, such as the role of parameters in statistics~\cite{cox1990role,box1979robustness} or assumptions in economics~\cite{spiegler2015behind}. One natural direction our exposition could therefore take would be to engage with this rich vein of existing theory as part of a comparative analysis linking these formalisms across domains. Although such a work would, we think, be an excellent contribution to the literature, we take a different and more prosaic tact to focus on the particular way in which HMAPs affect the practice of computing governance and research. But it is important to stress that our analysis is novel only in that specific focus and in our specific analyses that tread on distinct features of computation or on particular algorithmic models or constructions. There is, for certain, much to be gained in computer scientists' learning from the outcomes of hard-fought battles in related domains rather than relearning every lesson through bloody experience -- especially as subfields like cryptography and machine learning stumble into similar moral and practical questions previously faced by, \textit{e.g.}, environmental and atomic scientists. This is a particularly acute need for computing, given the way in which the products of the field continue to permeate throughout the sciences, humanities, civil society, and commerce, creating a broad spectrum of applications which must be governed.

\paragraph{Classifying HMAPs.} Our definition of HMAPs is not intended to be an exhaustive taxonomy of the obscure formalisms that can have outsized impact on the societal effects of computing. Such an exhaustive analysis could very well be of considerable value to the responsible computing community. However, undertaking such an extensive project within this paper would distract from our other goals of outlining the particular hazards such objects present and discussing ways in which the technical research community can work to better explain and motivate their importance and impact. In other words, our focus is more on developing a framework for recognizing and characterizing HMAPs that could then be applied more broadly, rather than on adding as many letters as we can to one particular acronym. Although this work deprioritizes complete classification, there are some distinguishing attributes of HMAPs (and other formalisms that are used in a similar way) that can be useful to keep in mind when considering their nature and implications. We will later link one useful attribute classification into our discussion of how the technical community can best communicate the nature of HMAPs, but, because it may be generally useful, we briefly highlight it here first.

HMAPs are used to formalize various social and physical processes and normative design decisions, and their selection and justification can stem from numerous different roots in the modeling process. For instance, the $\epsilon$ in differential privacy models a social concept (the individual privacy vs.~social utility tradeoff) in a manner where the efficacy of any given selection can be analyzed empirically -- as demonstrated by the whirling debate over its Census deployment. In general, HMAPs naturally fit into five different categories that capture what they model and how their use can be checked, addressed, or falsified: (i) \textit{formal}, (ii) \textit{formal-empirical}, (iii) \textit{physical} (which are inherently empirical), (iv) \textit{social-empirical} (like the $\epsilon$ of differential privacy), and (v) \textit{social}. We will often allude to this classification or explicitly state which category a particular HMAP fits into, and the distinctions among them often impact the nature of an HMAP and which hazards it presents.

\paragraph{Outline.} In~\S\ref{sec:hmaps} we detail these classes of objects and discuss their nature and their importance to the thorough analysis of computing and its harms. In~\S\ref{sec:six} we raise six hazards that make HMAPs uniquely treacherous for interdisciplinary analysis, demonstrated in~\S\ref{sec:ex} through more detailed discussions of post-quantum cryptography standardization, machine-learning hyperparameters, and, most extensively, differential privacy. We then consider how computer scientists might emphasize the nature and importance of HMAPs in a manner accessible to practitioners, the \emph{Systematization of Knowledge for Practitioners} (SoKfP), in~\S\ref{sec:pa}\cut{, before briefly concluding in~\S\ref{sec:conc}}{}.

\section{Heuristic Models, Assumptions, Parameters, and their Implications}\label{sec:hmaps}

\noindent It is a basic principle that the societal impact of a technology bears the imprint of the scientific process that developed it, and computing is certainly no different~\cite{cryptoeprint:2015:1162, mulligan2019thing, malik2020hierarchy, selbst2019fairness}. We are interested in artifacts that are the products of a particular pattern common in computer-science research: A social or physical problem is formally modeled, and an algorithm or protocol is developed that the researcher is \emph{just about} certain solves it. It may be that the algorithm works for a slightly different model, not equivalent to the original but `close enough' to seem justifiable. It may instead be that the algorithm seems to work and certainly does so if certain assumptions -- which are well defined on their own terms, independent of any particular application -- hold. Or, it may be that the algorithm produces an effect that could be desirable, but only if attuned to the circumstances of its use that are as yet unknown or are inherently subjective. So, the algorithm designer makes that effect adjustable and places the burden of selection onto the implementer, just as how the US Census Bureau was forced to determine an appropriate $\epsilon$ to apply differential privacy to disclosure avoidance. In any case, the algorithm or protocol is presented as a solution to the modeled problem, and the resultant artifact --  respectively, a heuristic model, assumption, or parameter -- is left as a consequential detail.

\paragraph{Heuristic Models.} Briefly, in this context by a \emph{model} we mean the system used to reason about the correctness of an algorithm or program. Most computational fields have, at least in some weak sense, a `standard model' that forms a baseline for analysis. For example, the standard model of cryptography considers an attacker with a classical (\textit{i.e.}, non-quantum) computer constrained only by limited time and computational resources.

There are two different forms heuristic models take, depending on whether our interest lies in mathematical or computational reasoning.\footnote{The term `heuristic model' invites an unfortunate clash of language, as the use of formal models and computation to inform mental or organizational reasoning is itself often a form of heuristic. Nonetheless, the usage of `heuristic' we invoke here is consistent with the language used in the computer-science literature.} For the former, a researcher is unable to prove some fact about an algorithm $A$ directly in the standard model $\mathcal{M}[A]$. So, the researcher instead modifies the algorithm $A \mapsto A'$ by replacing (a subroutine within) it with something simpler for which an argument can be articulated. This trick is then justified by introducing a heuristic model $\mathcal{M} \mapsto \mathcal{M'}$ which makes the modified $\mathcal{M}'[A']$ still `correct'. Non-standard cryptographic models are canonical examples, such as the random-oracle model (ROM)~\cite{bellare1993random} and the Fiat-Shamir heuristic~\cite{fiat1986prove}, both of which add additional assumptions onto the `standard model' in order to enable proofs. The use of these heuristic models can introduce additional complexities for analysis of protocols and their implementations. For example, Dao \textit{et al.}~and Khovratovich \textit{et al.}~each found that incorrect use of the Fiat-Shamir heuristic introduced attacks into cryptographic protocols used in various blockchain and verifiable computation applications~\cite{cryptoeprint:2023/691, cryptoeprint:2025/118}. For computational reasoning, the researcher has a program $P$ to be submitted as input to another program $\hat{P}(P)$ to be computed over. No such $\hat{P}$ is known or possible, but replacing $P$ with something simpler $P \mapsto P'$ makes $\hat{P}(P')$ practical. Program analysis often uses this technique, for example, when verifying security protocols -- such as with the Dolev-Yao model~\cite{barbosa2019sok, clarke2018handbook, dolev1983security}. In either case, the approximate nature of these models makes arguments relying upon them into only `heuristic proofs'~\cite{cramer1998practical}.

In both contexts, the suitability of the heuristic and the confidence derived from it relate to its fidelity, {\it i.e.}, the adequacy and accuracy with which $\mathcal{M}'$ represents the relevant structures of $\mathcal{M}$. However, even flawed heuristics may have value. The aforementioned ROM and Fiat-Shamir heuristic are both widely used within cryptography despite admitting security proofs for insecure constructions in certain pathological cases~\cite{goldwasser2003security, canetti2004random}. What makes a heuristic satisfactory cannot, by its nature, ever be formally settled. Acceptance is a social process within the technical community, ideally buttressed with formal analysis of evidence and implications. This process may be contentious, as demonstrated by the history of the random oracle model itself~\cite{canetti2004random, cryptoeprint:2006:461, koblitz2007another}.

\paragraph{Assumptions.} Presumption is inherent to scientific modeling as scientists propose theories to explain observations; thus, all but the most abstract computer science rests on uncertain beliefs somehow. Data sufficiency in machine learning, rationality in mechanism design, or adversarial modeling in information security are examples of modeling assumptions. Their plausibility cannot be divorced from the social or physical context of the problem, their validity determines whether a technically correct computation is practically useful, and their justification determines whether it can be socially beneficial. Thoughtful consideration of modeling assumptions, whether normative or positive, is perhaps the central focus of sociotechnical analysis of computation~(\emph{e.g.},~\cite{abelson2015keys, barocas2016big, buolamwini2018gender, dwork2019differential, gebru2018datasheets, kroll2016accountable, malik2020hierarchy, mitchell2019model, ohm2009broken, cryptoeprint:2015:1162, rudin2019stop, selbst2019fairness, weitzner2008information}).

Our interest instead lies with a distinct class of \emph{separable} assumptions, whose validity is instead independent of any specific practical application. These tend to be narrow statements that a specific mathematical or physical construction behaves as intended despite a lack of conclusive proof. Moreover, their validity tells only that a computation will in practice do what the modeling intends. The distinction in form between modeling and separable assumptions may be subtle and further complicated by how tightly coupled they often are. Given our claim that separable assumptions deserve explicit attention, it is important that we be able to distinguish them from modeling assumptions. We build towards a definition of separability as well as a qualitative distinguishing test between separable and non-separable assumption through examples, which also serve to demonstrate the importance of the former. Our definition centers on the ability of separable assumptions to stand on their own as statements whose validity (though not, necesssarily, their relevance) is independent of any particular application. Though they are far from the only `useful' assumptions in computing, we find separable assumptions to be far less commonly discussed as an essential element of problem and domain modeling than non-seperable ones.

One type of separable assumption are well studied yet unproven mathematical statements, believed true by researchers and used as if they are. Cryptography again provides us with canonical examples in the form of hardness assumptions~\cite{goldwasser2016cryptographic, naor2003cryptographic}, \emph{e.g.}, the RSA assumption~\cite{rivest1978method}, the Diffie-Hellman (DH) assumptions~\cite{diffie1976new}, and the Learning with Errors (LWE) assumption~\cite{regev2009lattices}. Because theoretical computer scientists do not yet have the tools to prove that there are no efficient algorithms for certain problems of interest in cryptography, many practical cryptographic constructions cannot be unconditionally proven secure~\cite{goldreich2009foundations, impagliazzo1995personal, katz2014introduction}. Most designs, including, \emph{e.g.}, all practical encryption schemes, instead rely on hardness assumptions, in the sense that there exists a rigorous proof that breaking the scheme requires invalidating the assumption. This allows study of the latter in isolation, and confidence in it justifies the claimed security.

As an example, when using the Decisional Diffie-Hellman (DDH) assumption, researchers assume that no practical adversary can determine whether a random element of some specific algebraic group $\mathbb{G}$ is independent of two others. The assumption directly states that no efficient algorithm can behave in a meaningfully different way when the element is independent compared to when it is not.
\begin{defi}[Decisional Diffie-Hellman (DDH) Assumption]\label{def:ddh}
Let $(\mathbb{G}, \, q, \, g) \leftarrow \mathcal{IG}(1^n)$ be an instance generator where $n, \, q \in \mathbb{N}$, and $g$ is a generator of group $\mathbb{G}$ of order $q$. For any probabilistic polynomial-time algorithm $\mathcal{A}$ and uniformly sampled $x, \, y, \, z \xleftarrow{\$} \mathbb{Z}_q$,
$$|\emph{Pr}[ \mathcal{A}(\mathbb{G}, \, q, \, g, \, g^x, \, g^y, \, g^z) = 1 ] - \emph{Pr}[ \mathcal{A}(\mathbb{G}, \, q, \, g, \, g^x, \, g^y, \, g^{xy}) = 1 ]| \leq \emph{\textsf{negl}}(n)$$
where $\emph{\textsf{negl}}(n)$ is eventually bounded above by the inverse of every polynomial function of the security parameter $n$.
\end{defi}

\noindent The truth of this statement for a specific $(\mathbb{G}, \, \mathcal{IG})$ domain does not depend on the context of its use, \emph{e.g.}, authenticity in end-to-end encrypted messaging~\cite{perrin2016xeddsa, melara2015coniks}. Notably, the technical requirement that $\mathcal{A}$ runs in "probabilistic polynomial-time" is a standard modeling assumption on adversarial capacity that underlies much of modern cryptography. This demonstrates how tightly coupled modeling and separable assumptions may be. While hardness assumptions are separable, they are only of interest because of an underlying modeling assumption about how capable adversaries are.

The DDH assumption has been carefully studied, and there are particular domains in which we are confident in its truth~\cite{boneh1998decision}. But it is not proven. A stroke of inspiration could find an $\mathcal{A}$ that violates the assumption on a domain used for a deployed cryptographic scheme and break it. Even the use of unrefuted assumptions may provide opportunities to a malicious actor. Cryptographic standardization requires the fixing of domains (and often specific instances) for schemes relying on hardness assumptions. Malicious generation can render those assumptions insufficient because of auxiliary information. Such a subversion almost certainly occurred with the DUAL\_EC\_DRBG cryptographically secure pseudorandom-number generator, which is widely believed to have contained a National Security Agency (NSA) backdoor~\cite{brown2007security, green2013many}. Even when researchers have a high degree of confidence in our assumptions, their use requires care.

Cryptography is far from the only source of separable assumptions in computer science. Of a similar `hardness' flavor are hardness (of approximation) statements like the bounds implied by the Exponential Time Hypothesis or the Unique Games Conjecture~\cite{impagliazzo2001complexity, khot2002power}. Much of the research investigating efficient (approximation) algorithms for important computational problems follows from the assumption by researchers that these results place hard bounds on the attainable quality of a solution. Although the incorrectness of any of these assumptions would not then invalidate any known algorithmic runtimes or approximation ratios, it would invalidate existing claims of conditional optimality that can discourage further attention from researchers. In an entirely different domain, another separable assumption arises in the use of Newton's Method in numerical analysis for scientific computing and optimization. The correctness and efficiency of Newton's Method is often dependent on assumptions on both the form of the function being approximated and the quality of the initialization state of the procedure, of which often neither can be reliably verified~\cite{boyd2004convex, laufer2023optimization}. Inevitably then, the practical use of Newton's Method rests on the separable assumption that these conditions are met and so the application of the algorithm is well founded.

As a further -- and distinctly different flavor -- of example, technical assumptions arise from executing programs on physical hardware that imperfectly implements a mathematical model of computation. It is inherent in all of computing, but especially acute in cryptographic engineering, formal verification, robotics, and cyberphysical systems, to assume processors, servos, and sensors appropriately transmute physical phenomena from and into mathematical information~\cite{smith1985limits}. Failures of these physical assumptions can cause great harm, \emph{e.g.}, the 2018-19 crashes of Boeing 737-MAX aircraft that killed 346 passengers. An essential element of these disasters was the behavior of control software when readings from sensors did not accurately represent the flight dynamics~\cite{ntsb}. At the intersection of cryptography and formal methods for program analysis and verification, side-channel attacks rely on the physically observable details of how specific software and hardware implement cryptographic schemes to steal private information by, \emph{e.g.}, exploiting data-dependent timing and cache usage~\cite{barbosa2019sok, percival2005cache}. Formal techniques to mitigate these attacks must assume that their protective measures will be rendered effectively when physically realized.

In each of these examples, the assumption is in some way separable from its potential uses. The DDH and other cryptographic hardness assumptions are well defined mathematical statements that would be valid objects of study even if they had no practical use in cryptography; whether Newton's Method converges is independent of the application of the function being approximated; a physical sensor on an aircraft is assumed to process information from the environment correctly, regardless of how that information is put to use for safe flight; general-purpose hardware is assumed to execute a program correctly, regardless of what the application of that program may be. We define separable assumptions as those that are both (i) \emph{concrete} and (ii) \emph{self-supporting}. Whether an assumption possesses these two attributes is therefore an affirmative (although only qualitative) distinguishing test.

For (i), a concrete assumption is of the form \emph{we assume this object has property $P$}, as opposed to generic assumptions of the form \emph{we assume there exists an object that has property $P$}. This small but important distinction is raised by Goldwasser and Kalai~\cite{goldwasser2016cryptographic} in the context of cryptography. Generic assumptions are speculative and lack a means for constructive use. Their proposal can alter the path of research, but only concrete assumptions impact the practical use of computation. All of our prior examples are concrete. The DDH assumption for a particular $(\mathbb{G}, \, \mathcal{IG})$ domain states that a specific problem is computationally hard. Physical assumptions are inherently concrete as they pertain to concrete implementations. In contrast, Goldwasser and Kalai provide examples of various generic cryptographic assumptions that we do not consider separable, such as the existence of one-way functions.

For (ii), a self-supporting assumption is both well defined and justifiable on its own terms, independent of how it is used. A researcher can reason about the correctness of the DDH without any reference to cryptography or evaluate whether a flight sensor measures accurately even if it is not connected to the rest of the avionics. In contrast, the validity of a modeling assumption -- the rationality of auction participants, the adversarial capacity available for attacking secure communications, the sufficiency of data to model an environment, \emph{etc.} -- is inherently tied to its deployment. It may always or never be well founded, or perhaps more likely will fall somewhere in between. But a modeling assumption can never be argued valid in general, independent of the particular context in which an algorithmic solution built upon it will be used. Simply put, modeling assumptions assert that we are solving the right problem, and separable assumptions imply that the solution to the problem as modeled is correct, even should that model prove useless.

\cut{An important corollary to self support is that separable assumptions can transfer to entirely unrelated contexts without any necessary reconsideration of their validity. If we somehow found a use for the DDH in machine learning, we could apply it with all confidence due to it from cryptography. In contrast, transferring notions of, \emph{e.g.}, rationality or adversarial intent to ML has required evaluating whether the agents in that setting possess analogous internal motivation and reasoning as in the economic and information-security contexts.}{}

\paragraph{Parameters.} Often the most conspicuously consequential of HMAPs, parameters allow researchers to concisely specify families of algorithms. Each family member has the same basic architecture, but members differ in function according to their identifying parameters. The choice of parameters then selects the family member most appropriate for a specific use case. This metatechnique allows computer scientists the flexibility to build expressive and generic theories, with nuanced application to an eccentric circumstance requiring only careful parameterization. However, it has a consequence. Parameters allow the reduction of social contention or physical uncertainty to numerical choice. Exemplified by an algorithm for fair ranking from the literature that we will shortly describe, it can be a deceptively simple choice at that. The sociotechnical implications of parameters are therefore often more immediate than for heuristic models or assumptions. Although they are not inherent to a modeled problem, parameters are frequently the intended means for its most stubbornly subjective qualities to be fixed into mathematical terms. An example of a consequential parameter choice is that of $\epsilon$ from differential privacy. Others appear in the explosion in technical enhancements for the beneficial use of machine-learning algorithms, \emph{e.g.}, fair and robust ML.

The goal of adversarially robust machine learning is to prevent an attacker given influence over the inputs to a machine-learned model from compromising its accuracy. A particular security concern is that an attacker may do so through carefully constructed perturbations indiscernable or unsuspicious to human or machine scrutiny~\cite{goodfellow2014explaining, madry2017towards, tsipras2018robustness}, \emph{e.g.}, with small stickers causing an autonomous vehicle to misread a stop sign~\cite{eykholt2018robust}. Tsipras~\emph{et al.}~\cite{tsipras2018robustness} model adversarial robustness as training a classifier with `low \emph{expected adversarial loss}'
$$\min_{\theta} \mathbb{E}_{(x,y) \sim \mathcal{D}} \bigg[ \max_{\delta \in \Delta} \mathcal{L}(x + \delta, \, y; \, \theta) \bigg]$$
for candidate model $\theta$, data point $(x, \, y)$ drawn from distribution $\mathcal{D}$, loss function $\mathcal{L}$, and $\Delta = \{ \delta \in \mathbb{R}^d \mid \| \delta \|_{p} \leq \epsilon \}$ a set of perturbations parameterized by $\epsilon$. Increasing $\epsilon$ enlarges this set and captures a strictly more powerful attacker. Its choice so fixes the maximal power of an adversary the learned model is trained to resist.

We might seem to therefore want as large an $\epsilon$ as possible. But, the effect of training against this loss is to produce a model reliant upon patterns in the data that are invariant when any perturbation $\delta \in \Delta$ is added to an input. This stability prevents those adversarial changes from altering the classification. However, it also compromises accuracy on fine distinctions where similarly small changes make all the difference between two inputs of different ground truth. Enlarging $\Delta$ through increasing $\epsilon$ extends this tradeoff to ever larger perturbations and coarser differences. Setting $\epsilon$ implicitly becomes a choice between prioritizing robustness and prioritizing accuracy. Beyond just this intuition, the authors of~\cite{tsipras2018robustness} are able to prove the existence of this tradeoff for a pathological distribution of specific structure. They also establish it experimentally, showing an inverse relationship between $\epsilon$ and accuracy on real datasets.

We must note Tsipras~\emph{et al.}~argue that this loss in accuracy may in fact `result in unexpected benefits: the representations learned by robust models tend to align better with salient data characteristics and human perception.' But both robustness and accuracy are desirable for the beneficial use of machine learning. Taken at face value, these results place them in conflict. Employing this approach to robust machine learning requires choosing through $\epsilon$ whether to accept a loss in accuracy for some security -- even when, in theory, that burden might fall disparately.

In the work of Celis~\emph{et al.}~\cite{celis2017ranking}, fairness in ranking problems, \emph{e.g.}, search-result prioritization, is modeled by placing lower and/or upper bounds on the number of entries with a given property that can appear at the top of the ranking. Although the algorithmic approach is generic, properties are naturally motivated through diverse representation. An implementation might, for example, require that images returned for search queries not over- or under-represent some social origin or character -- the present harms of which have been searingly analyzed and critiqued~\cite{noble2018algorithms}. The authors formalize this goal as the constrained maximization problem
$$\argmax_{x \in R_{m, n}} \sum_{i \in [m], j \in [n]} W_{ij}x_{ij}, \text{ such that }
L_{k\ell} \leq \sum_{1 \leq j \leq k} \sum_{i \in P_{\ell}} x_{ij} \leq U_{k\ell}, \, \forall \ell \in [p], \, k \in [n].$$
Here $R_{m,n}$ is the set of binary matrices indicating rankings of $m$ items into $n$ positions, $W_{ij}$ is the utility of placing item $i$ at position $j$ as determined by some arbitrary (and potentially biased) process, and for a property $\ell \in [p]$, $P_{\ell}$ is the set of items with that property. Most important for our consideration, the parameters $L_{k\ell}$ and $U_{k\ell}$ specify how many elements with property $\ell$ must or can be placed in the first $k$ entries. To instantiate a fair-ranking algorithm from this definition requires choice of these thresholds.

We must first consider whether it is appropriate to model fairness in a use of ranking through proportional representation with respect to a set of properties. If it is, then an implementer is still left with the choice of parameters that bound those proportions. This is a conceptually simple decision: `In the first $k$ results returned, no fewer than $L_{k\ell}$ and no more than $U_{k\ell}$ results may have property $p$' is easily understandable for a non-technical audience. However, there is no mathematical basis on which to make this choice. It is a subjective social and political question injected into the mathematical framework through parameterization, and any beneficial effect of adopting this algorithm is contingent upon it. While the function of these specific parameters are intuitive once that social character is recognized, the more sophisticated $\epsilon$ of differential privacy or adversarial robustness demonstrate that such simplicity cannot always be expected.

\cut{Another example of parameterization, in this case from outside of machine learning, arises in \textsc{Definition}~\ref{def:ddh}. The concrete security of a cryptographic scheme is how much computational effort is required of an attacker for some specified probability of a successful break. It allows estimation of the time and monetary investment an adversary may expect to spend directly attacking it, and it is how cryptographers ultimately interpret a `practical adversary.' Concrete security is tuned through choice of security parameter for the underlying hardness assumption(s). One such parameter is the $n$ of~\textsc{Definition}~\ref{def:ddh}. A major battlefield of the Crypto Wars of the 1990s was intentionally weakened export-grade cryptography. For the government of the United States, preventing the legal export of practically secure cryptographic systems was as simple as taking a theoretically secure design and mandating a security parameter (rendered through key length) with insufficient concrete security to prevent attacks~\cite{abelson2015keys, diffie2001export}. Even when we are confident in hardness assumptions and our reductions to them, concrete security is still contingent on careful parameterization.}{}

\vspace{1em}

\paragraph{Discussion.} Many computer scientists recognize that HMAPs can carry significant sociotechnical implications. One of the reasons for our frequent reference to cryptography is the seriousness with which its research community scrutinizes and accepts heuristic models, assumptions, and parameters, in order to robustly guarantee security and privacy (\emph{e.g.},~\cite{boneh1998decision, brown2007security, canetti2004random, cryptoeprint:2006:461, goldreich2009foundations, goldwasser2003security, goldwasser2016cryptographic, katz2014introduction, koblitz2007another, naor2003cryptographic, cryptoeprint:2015:1162,cryptoeprint:2023/691}). However, computer scientists possess the scientific background, mathematical maturity, accrued expertise, and, frankly, time, interest, and grant dollars to carefully consider the details of technical research. The rise in interdisciplinary collaboration and education -- joint degree programs, dual-track conferences and forums, and scholarship such as~\cite{ohm2009broken, barocas2016big, kroll2016accountable, malik2020hierarchy, mulligan2019thing, selbst2019fairness, weitzner2008information} -- is an encouraging sign that conspicuous sociotechnical concerns, such as modeling assumptions, will see sufficient consideration. For example, the quality of data used for machine learning has drawn commentary from numerous technical and humanistic perspectives, such as in~\cite{barocas2016big, buolamwini2018gender, gebru2018datasheets, malik2020hierarchy, mitchell2019model, mulligan2019thing, noble2018algorithms, selbst2019fairness}. However, broad and thorough treatment of the obscurer HMAPs is more irregular.

Three influential articles demonstrate a lack of explicit consideration of HMAPs at the intersection of computer science and law: Ohm's \emph{Broken Promises of Privacy}~\cite{ohm2009broken}, Barocas and Selbst's \emph{Big Data's Disparate Impact}~\cite{barocas2016big}, and Kroll~\emph{et al.}'s \emph{Accountable Algorithms}~\cite{kroll2016accountable}. All three (excellent) articles were written by technically adept and knowledgeable authors. Those of Kroll~\emph{et al.} include multiple active computer-science researchers. Each article provides detailed, thorough, and accessible analysis of the key modeling questions and proposed technical designs within their scope. However, their treatments of HMAPs are limited, and therefore many of the algorithms and techniques they present as valuable governance tools are analyzed without consideration of their fundamentally qualified and conditional nature. Ohm only reproduces a figure from~\cite{brickell2008cost} demonstrating that a privacy vs.~utility tradeoff at the heart of his analysis can be parameterization-dependent but otherwise leaves `the full details... beyond the scope of this Article.' Barocas and Selbst quote that Dwork~\emph{et al.}~`demonstrate a quanti[t]ative trade-off between fairness and utility' in their influential \emph{Fairness through Awareness}~\cite{dwork2012fairness}. No mention is made, however, of the technical conditions under which the tradeoff is provable, nor that Dwork~\emph{et al.}~propose for scientists and engineers to tune it through parameterizations for constraining bias and the learning of metrics. Meanwhile, Kroll~\emph{et al.}~identify multiple areas that depend on HMAPs: program analysis, cryptographic commitments, and zero-knowledge proofs. Nonetheless, their consideration of it is limited and indirect, comprised mostly of discussion about how accountable randomness requires addressing assumptions that theorists usually make about its physical collection.

This is not surprising. These articles focus on the law, not the technology. The audience is composed of lawyers who may not have the background for or interest in any further detail, and ample references guide those who do. The choice of how much technical detail to incorporate was, we are sure, a thoughtful, measured, and ultimately wise and good decision on the part of all the authors. But it is the clear intent of each of these articles to be a reference and resource, to help guide and shape how the scientific elements of difficult sociotechnical questions raised by privacy, accountability, and data mining are discussed and understood in legal scholarship and practice. Just because detail is withheld for good pedagogy does not mean that detail is irrelevant, and the understanding of practitioners must incorporate the implications. HMAPs often fall into a grey zone. Their exclusion from foundational work at the intersection of computer science and law is in some sense justified by their obscurity and opacity, but it remains consequential to the ideas present in that scholarship.

For example, in evaluating barriers to reform of social harms from computational data analysis, Barocas and Selbst write `[s]olutions that reduce the accuracy of decisions to minimize the disparate impact caused by [data modeling] will force analysts to make difficult and legally contestable trade-offs.' The fair-ranking scheme of Celis~\emph{et al.} demonstrates the importance of HMAPs to such choices. The base unconstrained maximization problem requires no compromise, so the `legally contestable' nature of the enhanced algorithm derives entirely from `$\text{such that } L_{k\ell} \leq \sum_{1 \leq j \leq k} \sum_{i \in P_{\ell}} x_{ij} \leq U_{k\ell}, \, \forall \ell \in [p], \, k \in [n].$' Any question of its acceptability and legality would reduce to (i) the modeling decision to introduce that constraint schema, (ii) the further modeling decisions of choosing properties and classifying items to define the $P_{\ell}$ sets, and (iii) the choice of parameters $L_{k\ell}$ and $U_{k\ell}$. The full burden of `fairness' falls upon the validity of all three. It is to the great credit of Celis~\emph{et al.}~that $L_{k\ell}$ and $U_{k\ell}$ have such simple and approachable mathematical function. But not all fairness parameters may be so explicable, let alone all HMAPs of sociotechnical importance. Establishing the full measure of the type of solutions that Barocas and Selbst propose will require pushing past the basic principles of formal representations of 'fairness' into some of the more obscure mathematical details.

Without care, consequential questions around HMAPs may come to be viewed outside of computer science as mere implementation details, especially when they are inscrutable to non-technical observers. If so, the heavy burden and significant influence those questions can bear may be delegated to people and processes without the insight to act effectively in the public interest. Moreover, the various academic, ethical, legal, political, and societal mechanisms that govern computing will fail to capture the full challenge faced by those who manage these technologies in practice. They will be rendered at best incomplete, at worst impotent.

\section{Six Hazards}\label{sec:six}

To this point, we have justified our interest in HMAPs through reference to how they can bear significant risk while appearing obscure, opaque, inconspicuous, or pedantic. However, HMAPs possess other qualities that intensify the challenge to sociotechnical scrutiny they present. We see HMAPs as presenting six hazards to broad and thorough analysis -- in a manner notably distinct from modeling choices and algorithm design -- especially for practitioners with limited scientific expertise. We also briefly note that, like HMAPs themselves, these hazards overlap significantly with veins of both philosophical theory and analysis of existing practice in adjacent domains, like statistics and economics~\cite{cox1990role, box1979robustness, spiegler2015behind}.

\paragraph{Hazard \#1: Obscurity and Opacity.} Explicitly, \emph{while their uses may have significant social consequences, HMAPs are themselves of a purely technical nature. So, their existence or basic function may not be apparent to those without non-trivial scientific understanding.}

\paragraph{Hazard \#2: The Pretense of Formalism.} That first hazard is only compounded by how \emph{HMAPs often carry a pretense of formalism} when of mathematical origin. While algorithms may be presented in mathematical terms, our common language captures their creative dimension. We \emph{design} algorithms, we \emph{write} programs, we \emph{construct} protocols. They do not bring the sense of inevitability that HMAPs may. Heuristic models, assumptions, and parameters often present as mathematical detail and so as inherently true or false, right or wrong, in a way the subjective or empirical nature of modeling does not. Some of these objects do have well defined truth values. Cryptographic hardness assumptions, like the DDH over a given $(\mathbb{G}, \, \mathcal{IG})$ domain from~\S\ref{sec:hmaps}, are well defined mathematical statements. We just have not (yet?) firmly established their truth~\cite{boneh1998decision, goldreich2009foundations}. Others, especially heuristic models including, \emph{e.g.}, the aforementioned ROM~\cite{bellare1993random}, are often not `provable' because of theoretical technicalities. Any impression they give of formal truth is likely harmless outside scientific discourse. But many other HMAPs -- especially parameters -- often have only a social or empirical basis. Their presentation as mathematical ideas and notation, \emph{e.g.}, the $\epsilon$ of adversarial robustness or differential privacy, may inadvertently launder that amathematical character.

\paragraph{Hazard \#3: Inaccessibility.} Technical simplicity can aid scrutiny of the sociotechnical repercussions of HMAPs. The simple function of $L_{k\ell}$ and $U_{k\ell}$ is, \emph{e.g.}, a great strength of the fair-ranking approach of Celis~\emph{et al.}~\cite{celis2017ranking} as discussed in~\S\ref{sec:hmaps}. However, in general such clarity is never assured, highlighting how \emph{HMAPs that demand careful sociotechnical scrutiny may be technically complex and thus inaccessible to practitioners.} The $L_{k\ell}$ and $U_{k\ell}$ parameters avoid this hazard. Although they carry immense social depth, once that is recognized it is not hard to imagine reasonable processes -- commercial, legislative, regulatory, judicial -- for choosing them. They are simple enough in technical form and function to be consistent with traditional modes of resolving social contention. In contrast, for adversarial robustness the simple structure of $\epsilon \in \mathbb{R}$ paired with its immense impact makes it far harder to conceive of a selection method able to comprehensively translate its sociotechnical dimensions to and from its mathematical choice. Even the smallest tweak can lead to new inaccuracies for a machine-learning model with significant social consequences, even if on the whole the test accuracy of the model is barely degraded. This difference stems from the gap between the focused and transparent operation of the fairness parameters and the sweeping fullness of $\epsilon$. While the former have a clear and self-contained function as targeted and modular components of a broader ranking system, the latter bears a much greater sociotechnical load as the unique fulcrum balancing robustness against accuracy.

\paragraph{Hazard \#4: Indeterminism and Unfalsifiability.} Each of the preceding hazards concerns whether the nuances of HMAPs will be adequately conveyed to practitioners. The fourth hazard works in the opposite direction -- it speaks to whether computer scientists can assuage concerns raised when \emph{HMAPs are indeterminate or unfalsifiable}. By \emph{indeterminate}, we mean that there is no technical notion that captures all of the failures that can result from their use. By \emph{unfalsifiable}, we mean there is no (practical) mechanism by which its choice or acceptance may be shown incorrect on purely technical grounds. In other words, we cannot know every way indeterminate HMAPs can fail, while we are unable to demonstrate that the use or choice of unfalsifiable HMAPs must be wrong. Indeterminate and unfalsifiable HMAPs can complicate demonstration of the (non-)existence of risk and harm from computing, which is essential for its thorough sociotechnical analysis. Although this hazard is less distinctive of HMAPs, it takes on a harder edge in the shadow of its first three siblings.

Heuristic models and assumptions may be given a formal structure independent of context and so are most often determinate either formally or empirically. We can know, at least in principle, what their failures could be and the technical consequences. All heuristic models are unfalsifiable by definition. The ROM and Fiat-Shamir heuristics are as noted provably false in general~\cite{goldwasser2003security, canetti2004random}, which is why reliance upon them will always be uncertain. Assumptions are usually falsifiable. Although not able to be `disproven' mathematically, most physical assumptions relevant to deployed computing may be shown invalid in relevant contexts through robust empirical investigation by engineers and natural scientists. Mathematical assumptions are (up to the independence phenomenon) falsifiable by rigorous proof.

Indeterminate and unfalsifiable parameters are pervasive. For many, the subjective nature of their selection precludes any purely technical notions of falsity or failure. The $\epsilon$ of adversarial robustness, the $\epsilon$ of differential privacy, and the $L_{k\ell}$ and $U_{k\ell}$ of ranking fairness all are so. There is no inherently right or wrong choice of tradeoff between robustness vs.~accuracy, privacy vs.~utility, or fairness vs.~(perceived) utility respectively, except under a subjective and prescriptive sociotechnical judgement. It is worth noting, however, that other parameterizations may be placed on mathematical foundations that allow empirical analysis of their validity. Not all are indeterminate or unfalsifiable. Tuning the concrete security of a cryptographic scheme through choice of $n$ is an excellent counterexample~\cite{alagic2022status, boudot2020factoring, chang2012third}. Some parameterizations of a social or physical character may also have determinate and falsifiable implications. For example, in the theory of machine learning, a parameterized choice of desired accuracy and confidence provably dictates the theoretically required size of the training dataset, known as the sample complexity~\cite{shalev2014understanding}.

\paragraph{Hazard \#5: Research Attention.} The final two hazards relate to how the broader systems built around computation manage HMAPs. One concern is that \emph{study of HMAPs may have a limited constituency within the research community}\ --\ or, at least, a limited constituency in comparison to the size of the community for the relevant theory\ --\ not all researchers in which may be interested in investigating such detail. An important variant of this hazard is when interdisciplinary evaluation of HMAPs is required by its social or physical character and implications, but little collaboration arises. For example, and as we will discuss in further detail in \S\ref{subsec:dp}, attention from economists, social scientists, and statisticians on parameterization of differential privacy has -- despite some excellent work -- been dwarfed by interest in its theoretical refinement and extension~\cite{abowd2019economic, desfontaines2020sok, dwork2014algorithmic, garfinkel2018issues, lee2011much, hsu2014differential, wasserman2010statistical}.

Computer-science researchers may believe that it is the responsibility of practitioners to do the hard applied work of developing parameterizations. Even if this particular burden shifting does not transpire, analyzing parameter choices and the reliance of constructions on heuristic models and assumptions may be seen as a secondary effort for which researchers will not be as well rewarded professionally compared to developing new theoretical results. Although far from perfect, cryptography remains the gold standard on addressing these hazards and provides an example of valuing such research efforts. Hardness assumptions often receive substantial, independent attention from leading researchers~\cite{boneh1998decision, regev2009lattices}, and a small number of cryptography researchers devote a considerable portion of their research programs to keeping the use of many physical assumptions and parameters honest, \emph{e.g.}~\cite{bernstein2013factoring, boudot2020factoring, halderman2009lest, heninger2012mining}. This line of applied analysis has been essential for confidence in the security of deployed cryptography.

\paragraph{Hazard \#6: A Soft Underbelly.} Finally, \emph{HMAPs may form a soft underbelly in regulatory and legal mechanisms, as well as in the positive use of computation to address social problems}. Any effort to manage technology through qualitative evaluation, mandates, and procedures must be sensitive to its technical eccentricities. The examples of backdoored standards and export-grade restrictions in cryptography (see~\cite{abelson2015keys, brown2007security, diffie2001export} as raised in~\S\ref{sec:hmaps}) show how HMAPs may be vehicles for politically motivated compromise. The converse -- where the subjective or indeterminate elements of a technology are used to subvert the governing infrastructure built around it -- is also of grave concern.

Without thoughtful analysis stimulated by deep technical understanding of HMAPs, their careless, negligent, or malicious use might defang regulation and legislation. The assurance intended by a requirement for formal verification of a cyberphysical system depends entirely upon the validity of the physical assumptions underlying that analysis. Should a search engine be required to integrate algorithmic techniques to minimize bias, its engineers might reach for Celis~\emph{et al.}'s fair-ranking algorithm, with the resultant effect fully at the mercy of choice of $L_{k\ell}$ and $U_{k\ell}$. Mandates both for machine-learning accuracy and for its robustness, if the latter were instantiated through an approach like that of Tsipras~\emph{et al.}, would be in direct tension, only resolvable through the contentious selection of $\epsilon$. To the extent that ideal parameters can exist in some qualitative and subjective sense, they may be application-specific and not amenable to global standardization as cryptographic key lengths are. Any legal finding of fact as to whether a choice of $L_{k\ell}$ or $\epsilon$ produced harm would require significant technical expertise. Proactive regulatory approval would further require availability of that expertise on a considerable scale, given the growing use of ML. Such problems will only proliferate alongside the ever increasing use of computation.

In general, without public infrastructure conscious of the need to evaluate and guide dependence on HMAPs, it may be difficult to bring about desired social effects. Further, the nuances of HMAPs are shaped by modeling; so they are only addressable after clearing a first hurdle of designing legal and regulatory systems around the intricate difficulties of the latter. The muddier that process is, the harder establishing consensus around HMAPs may very well be. Uncertainty as to whether and how robustness should be mandated for machine learning, \emph{e.g.}, might complicate the ultimate selection of $\epsilon$ as stakeholders disagree on appropriate burdens and goals.

The potential of HMAPs to subvert oversight also applies to any suggestion, like that of Abebe~\emph{et al.}~\cite{abebe2020roles}, to use computation as a tool for positive social change. As they write, `because they must be explicitly specified and precisely formalized, algorithms may help to lay bare the stakes of decision making and may give people an opportunity to directly confront and contest the values these systems encode.' However, the uncertain and subjective nature of HMAPs strikes at the ability to understand just what exactly systems built on computation encode and express. As a general principle, this final hazard states that, when the effect of a computation is dependent upon HMAPs, any attempt to control and wield it for social gain is reliant upon due consideration of its heuristic models, assumptions, and parameters. This is particularly precarious in light of the first five hazards.

\section{Examples}\label{sec:ex}

We give more in-depth consideration to three indicative examples of the influence of HMAPs. Each instance is integral to a topic that is currently a focus of sociotechnical analysis and debate.

\subsection{Post-Quantum Cryptography Standardization}\label{subsec:pqc}

Many popular hardness assumptions that underlie widely used cryptographic schemes have a curious status. Although for the moment intact in the face of classical adversaries, they are known to be broken against an attacker with an (as yet hypothetical) quantum computer of sufficient capability. Examples include the RSA~\cite{rivest1978method} and Diffie-Hellman (DH) assumptions~\cite{diffie1976new}, which are theoretically susceptible to the quantum polynomial-time algorithms of Shor~\cite{shor1999polynomial}. A quantum computer capable of running Shor's algorithms on real-world cryptographic deployments may be decades or more away. Nonetheless, some actors -- particular intelligence and defense agencies -- have need for even longer assurances of information security, especially in the face of `harvest now, decrypt later' threats. As such, in 2022 the United States government, through the National Institute of Standards and Technology (NIST), standardized a first set of post-quantum cryptographic (PQC) schemes based on assumptions believed secure against even quantum adversaries~\cite{alagic2022status}. Given the influence of prior NIST standardization processes, it is very possible that in the near future such algorithms will be required to be supported by any cryptographic technology used -- by rule -- by the United States government and -- by convention -- most global commerce.

With one exception the standardized algorithms are all instances of \emph{lattice-based cryptography} and derive their security claims from the hardness of (variants of) the LWE problem~\cite{regev2009lattices}. However, the standardization process received numerous submissions reliant on many alternative mathematical assumptions, some of which were found to collapse under the close scrutiny placed on them during the highly competitive process. In multiple cases, proposed schemes were found to be quickly breakable by classical computers. In other cases, assumptions were not found to be broken \emph{per se} but were shown to be much weaker than originally believed, as new attacks chipped away at the concrete security provided. This led to adjustments in which the parameterizations of the submissions were tweaked to increase the margin of security provided by the schemes. This arms race was very much in the spirit of the open competition fostered by NIST, in which teams (and external researchers) were invited to analyze and attack the constructions of their competitors. However, it laid bare how many of the new assumptions, based in many cases on much more complex mathematical objects than the RSA or DH problems, were not as well understood as the community may have hoped.

Take, for instance, the following account of how the submission of Rainbow, an instance of \emph{multivariate cryptography}, was tweaked during the standardization process~\cite{beullens2022breaking}. The specific details of the scheme, its parameters, and the attacks are beyond our scope. Relevant to our analysis is instead the way parameterizations were used to paper over weaknesses in the concrete hardness of the relied upon theoretical assumption (citations removed):
\begin{quote}
  \small
The participation
of Rainbow in the NIST PQC project motivated more cryptanalysis. During the
second round of the NIST project, Bardet \emph{et al}. proposed a new algorithm for
solving the MinRank problem. This drastically improved the efficiency of the MinRank attack, although not enough to threaten the parameters submitted to NIST. A more memory-friendly version of this algorithm was proposed by
Baena \emph{et al.} Perlner and Smith-Tone tightened the analysis of the Rainbow Band Separation attack, showing that the attack was more efficient than previously assumed. This prompted the Rainbow team to increase the parameters slightly for the third round. During the third round, Beullens introduced new attacks which reduced the security level of Rainbow by a factor of 220 for the
SL 1 parameters. The Rainbow team argued that, despite the new attacks, the
Rainbow parameters still meet the NIST requirements.
\end{quote}
Unfortunately for Rainbow this back-and-forth was not the end of the story. This excerpt is drawn from the introduction of Beullens' \emph{Breaking Rainbow Takes a Weekend on a Laptop}~\cite{beullens2022breaking}, which showed that the scheme as then proposed was still susceptible to a classical attack that could, in fact, be executed over a weekend (\textasciitilde 53hrs) on a commodity laptop. Although Beullens noted that yet another adjustment to the parameterization might limit the practicality of the attack, he also noted that the attack itself could be strengthened further. In the aftermath of this attack -- and the aspersions cast on the assumptions underlying Rainbow in general -- it was not chosen for standardization. Nor would any other multivariate cryptographic construction be selected either.

Another example occurred with Supersingular Isogeny Key Encapsulation (SIKE), based on Supersingular Isogeny Diffie-Hellman (SIDH) and the standard bearer for \emph{isogeny-based cryptography} more generally. Although not selected for standardization, SIKE was originally selected for continuation into an (ongoing) fourth round of the NIST competition~\cite{alagic2022status}. Shortly afterwards, Castryck and Decru announced an attack on SIDH~\cite{castryck2023efficient} that was able to break the proposed SIKE parameters within a day on a laptop and then was quickly further improved to get the time to under an hour~\cite{oudompheng2022note}. Their attack was based on applying a decades-old result from algebraic geometry whose efficacy against SIKE and SIDH had not yet been fully harnessed in the prior cryptanalysis literature. In addition to undercutting the NIST submission, Castryck and Decru were also able to claim a \$50k USD prize put up by Microsoft (where part of the SIKE team was based) as part of a challenge to motivate external cryptanalytic attention on the scheme~\cite{msSIKE}.

Even the lattice-based schemes ultimately selected were not without controversy. Numerous mailing-list fights and some publications were devoted to arguments about their security and parameterization, many of which were rife with personal animosity and accusations of motivated reasoning among the participating researchers~\cite{cryptoeprint:2019:691, emails}.

\paragraph{The Six Hazards.} The post-quantum cryptography standardization process makes for an excellent example of the research community's navigating the six hazards not just despite -- but in a sense because of -- the existence of Rainbow, SIKE, and other flawed submissions. A robust process encourages exploration held up to tough but fair scrutiny.

Cryptographic assumptions and the resultant theoretical and concrete security of constructions based on them are formal objects, but they are certainly obscure, opaque, and inaccessible. The process of standardization itself is designed to mitigate these three hazards, by having a small group of computer scientists reach a consensus on what is secure that the rest of the world can then blindly follow. This process takes advantage of an exceptionally rare quality nonetheless present in modern cryptography: With due care, researchers can make global choices of assumptions, algorithms, and parameters tuned against conservative predictions of present and future adversaries up to and including nation states and quantum computer-wielding attackers. The choice of schemes and parameters to standardize is empirical (albeit supported by strong theoretical underpinnings), but it is not social. There can in principle be one `good enough' answer for everybody. Of course, as the history of the DUAL\_EC\_DRBG pseudorandom-number generator backdoor demonstrates, a choice of cryptographic standard may in fact be `good for me, but not for thee.' However, the open nature of the PQC process, as well as its international, academic, and commercial participation, renders direct backdoors unlikely. The intended future use of the standards by the United States' own security community provides even greater confidence that the standards are not a soft underbelly to be intentionally subverted.

Nonetheless, the other hazards render the process still fraught. Although choice of assumptions is falsifiable and selection of parameters can, with care, be determinate, at the cutting edge there is still an immense amount of fallible guesswork. Heavily scrutinized, justified, and debated guesswork, but still, in the end, guesswork, as demonstrated by the near misses of Rainbow and SIKE. Moreover, although the PQC standardization process provided a forum for significant and successful research attention on the HMAPs relevant to its candidates, the length and depth of the scrutiny is rare even by cryptographic standards. The example of SIKE, which had already seen significant research attention before the competition, lays bare how much scrutiny can be needed for a design to truly be trustworthy.

\subsection{Machine-Learning Hyperparameters}\label{subsec:mlh}

Few if any computing technologies are at present under as extreme scrutiny from legal and policy experts -- and society at large -- as machine learning. The autonomous or semi-autonomous use of automated decision making and generative models brings significant promise to applications throughout nearly every facet of modern society. But it also brings a commensurate risk of social harms and economic upheaval. At present lawmakers, policymakers, and regulators are scrambling to catch up to the explosion in both hype and deployment of ML-based systems -- and generative large language models (LLMs) in particular. Just at the federal level of the United States government over a little more than six month period bridging late-2022 into early-2023, the White House Office of Science and Technology Policy proposed a Blueprint for an AI Bill of Rights~\cite{ostp}, the Federal Trade Commission (FTC) and Department of Justice (DOJ) joined with other regulatory agencies to put out a joint statement reiterating their commitment to actively regulate AI~\cite{js}, and Congress held widely covered hearings on the societal and policy implications of these new technologies~\cite{congress}.

At the same time, research into machine learning and its many applications is facing a burgeoning reproducibility crisis~\cite{kapoor2022leakage}. Although there is a robust theoretical framework for ML~\cite{shalev2014understanding}, our practical understanding of these models is most often driven by empirical data and -- especially in the case of generative tools -- anecdotal experience. There are a number of different mechanisms that can render the results of an analysis of an ML system suspect (see~\cite{kapoor2022leakage} for a survey). One particular source of difficulty can be underspecified or overtuned choices of \emph{hyperparameters}~\cite{cooper2021hyperparameter}.

By convention, the parameters of an ML model are its internal constants derived algorithmically during its learning phase, and then used to evaluate new inputs in deployment during the inference phase. Our `parameters', \emph{i.e.}, in the sense of HMAPs, are instead referred to as hyperparameters (or sometimes \emph{metaparameters} or \emph{tuning parameters}). Hyperparameters range from objects of obvious social importance -- such as the examples of fairness (hyper)parameters previously discussed -- to much more obscure technical objects. For example, a canonical hyperparameter in machine learning is the \emph{learning rate} (often notated $\alpha$), which is a scalar factor that affects how much the model is tweaked in response to mistakes made by it on the training data during the learning phase. Another example, particularly relevant to generative AI, is the \emph{temperature}. Many generative and classification models produce not only a distinct output, but a probability distribution over many possible outputs -- which doubles as both a source of alternative assessments and a measure of confidence in that singular output. The temperature adjusts how confident the model should be in finding a unique judgement, against `flattening' the probability distribution to put more weight on alternatives~\cite{hinton2015distilling}. Choice of hyperparameters has an immense effect on the accuracy of trained models, to the point that hyperparameter optimization (HPO) has become its own distinct field of study within the ML community~\cite{feurer2019hyperparameter}.

Many common machine-learning hyperparameters like the learning rate and temperature have a distinct nature as either \textit{social-empirical} or $\textit{physical}$ objects -- respectively depending on whether the process being learned is social or physical in nature. In some sense, they are poor examples of HMAPs, because they are often set purely empirically during training by finding which selection leads to the best results on benchmarks. As such, at first glance, there can seem to be nothing particularly impactful about them beyond their influence on those benchmarks; therefore, they can seem to be of little social importance. However, the straight line view of machine learning -- modeling to data harvesting and preprocessing to learning to validation to deployment -- is increasingly being replaced by viewing ML-based systems as forming lifecycles~\cite{gebru2021datasheets, nickel2024no}. Almost inevitably, the design and deployment of ML-based systems turns out to be an iterative process in which the efficacy of models generated during previous cycles informs changes and improvements to the process for the next iteration. In the lifecycle model, the empirical selection of such hyperparameters begins to take on a more dynamic and socially relevant role, because the decision by the implementer to accept a certain level of performance as attainable through hyperparameter optimization may drive how they choose to update their data, model architecture, and benchmarks in future cycles.

Moreover, other hyperparameters more clearly fit our the mold of HMAPs as objects that shift responsibility for normative design decisions onto the implementer. For instance, many ML classification models work by converting a confidence score into a prediction using a threshold hyperparameter. In the case of, say, a binary classification model prescreening resumes for hiring, choice of the threshold exactly determines which candidates move on to a human assessment.

Hyperparameters present a number of challenges for interpretation and replication of ML-based research. One major problem is overtuning, where researchers find hyperparameters than are optimal for their benchmarks but may not generalize, making their results of limited value. This may make it hard for governance bodies to assess the true state-of-the-art for various tasks and overall capability of ML methods, as well as to determine the effectiveness of technical `mitigations' such as fairnness metrics and constraints. Cooper \emph{et al.} study this problem with an eye towards a formal framework that aims to prevent this `gaming' of empirical ML studies~\cite{cooper2021hyperparameter}. Another concern for open and comprehensive assessment of ML claims is that hyperparameters and other model properties are increasingly kept private in the name of intellectual property and propriety. This secrecy is inexorably tied to concerns about centralization of ML technologies and their monopolization and regulatory capture by the organizations that create them. The necessity for secrecy is sometimes even justified by as yet unfounded claims of existential risk due to a `runaway' artificial general intelligence (AGI)~\cite{congress}, as occurred with the release of GPT-4 by OpenAI, who refused to disclose many critical technical details about their hyperparameters, architecture, and overall method~\cite{openai2023gpt4}.

\paragraph{The Six Hazards.} Machine-learning hyperparameters receive intense research attention. However, they test the five remaining hazards. Hyperparameters certainly can be obscure and opaque. Many refer to particular technical details of not just machine-learning models in general like the learning rate, but even specific neural network or other model architectures. Hyperparameters can also carry the pretense of formalism, as many can be analyzed through computational learning and other algorithmic theories, albeit in a way that does not necessarily fully capture their empirical behavior~\cite{cohn1990can, cohn1992tight}. The behavior of the parameters themselves can also lead them to be inaccessible. Some, such as the learning rate, sample complexity, or $L_{k\ell}$ and $U_{k\ell}$ of Celis \textit{et al.}, are not overly difficult to convey, hence our invocation of them in this article. However, others, such as the hyperparameters that define convolutional architectures so essential to effective image models, can be far more mathematically intensive to understand the implications of.

Hyperparameters are almost always indeterminate, being of a social or empirical nature, at least when it comes to their practical effect. Once again, fairness (hyper)parameters are a canonical example, as are classification thresholds, but even acceptance of a parameter like the learning rate often comes down to an informal determination that it is `good enough' based on empirical validation that may or may not match the real-world distributions. When falsifiable, they are so through computational learning theory, which may or may not be directly applicable to the empirical practice of machine learning itself. Hyperparameters can also certainly provide a soft underbelly for subverting governance of machine learning, because, when indeterminate or not strongly specified, the ultimate societal effects of an ML deployment can have immense variation.

\subsection{Differential Privacy}\label{subsec:dp}

We last return to a parameterization that exemplifies all of our hazards, the $\epsilon$ of differential privacy~\cite{dwork2006calibrating, dwork2008differential, dwork2009differential, dwork2014algorithmic, wood2017differential}. DP provides a rigorous and provable definition of privacy for statistical database queries, as heuristic anonymization techniques are often susceptible to attacks~\cite{dinur2003revealing, narayanan2008robust, ohm2009broken}. The principle of DP is to associate privacy not with constraining data collection or collation, but rather data analysis.

DP is an \emph{indistinguishability} definition derived from cryptographic theory~\cite{goldwasser1984probabilistic}. Informally, two probability distributions are indistinguishable if no adversary can determine from which (a sequence of) observed samples are drawn. \textsc{Definition}~\ref{def:ddh} is an example. This fundamental concept extends to one of \emph{near indistinguishability}, where nearness is moderated through a parameter $\epsilon$. The justification for DP is that, \emph{if the outcome of a statistical analysis is near-indistinguishable between databases that differ only in the inclusion of a single entry, then an adversary cannot learn too much about that entry from the outcome of the analysis alone}. A differentially private \emph{mechanism} is an algorithm for answering statistical queries guaranteed (usually through the introduction of calibrated noise) to be insensitive to any specific data entry, yet still give a reasonable approximation of the desired statistic. The original formulation of DP makes near indistinguishability rigorous through the following parameterized definition~\cite{dwork2006calibrating}.
\begin{defi}[$\epsilon$-indistinguishability]\label{def:ei}
  Two random variables $A$ and $B$ are $\epsilon$-indistinguishable, denoted $A \approx_{\epsilon} B$, if, for all measurable sets $X \in \mathcal{F}$ of possible events:
  $$\emph{Pr}[A \in X] \leq e^{\epsilon} \cdot \emph{Pr}[B \in X] \text{ and } \emph{Pr}[B \in X] \leq e^{\epsilon} \cdot \emph{Pr}[A \in X].$$
\end{defi}
\noindent To provide a formal theory of data privacy, DP operates over a universe $\mathcal{D}^n$ of databases with $n$ entries drawn from some domain, such as $\mathbb{R}^d$. For a given database $D \in \mathcal{D}^n$, an adjacent database $D^{-}$ differs from $D$ only in having one constituent entry deleted. A mechanism $\mathcal{M}$ is then identified with a random variable taking on its output when executed over a given database.
\begin{defi}[$\epsilon$-differential privacy]\label{def:dp}
  A mechanism $\mathcal{M}$ is $\epsilon$-differentially private if, for all adjacent databases $D \in \mathcal{D}^n$ and $D^{-} \in \mathcal{D}^{n-1}$, $\mathcal{M}(D) \approx_{\epsilon} \mathcal{M}(D^{-}).$
\end{defi}
\noindent This definition is not unique. According to a 2020 survey, `approximately \emph{225} different notions, inspired by DP, were defined in the last 15 years'~\cite{desfontaines2020sok}. Each variant modifies one or more aspects of DP through an alternative mathematical formulation. Many of these definitions require distinct parameterizations, and some have been shown to introduce vulnerabilities to reconstruction attacks of exactly the kind that DP was invented to prevent~\cite{protivash2022reconstruction}.

Any theory of privacy built on \textsc{Definition}~\ref{def:ei} depends absolutely on careful choice of $\epsilon$, demonstrated in the extreme by the fact that any two distributions on the same support are $\epsilon$-indistinguishable for all
$$\epsilon \geq \sup_{X \in \mathcal{F}} \bigg| \ln \bigg(\frac{\text{Pr}[ A \in X ]}{\text{Pr}[ B \in X ]} \bigg) \bigg|.$$
If, \emph{e.g.}, we let $A$ take on a fair coin flip and $B$ take on a biased coin flip with a 90\% chance of heads -- two distributions hardly indistinguishable in any intuitive sense of the word -- then $A \approx_{1.61} B$.\footnote{This example, while simple, is not idle. Coin flips are the basis of \emph{randomized response}, a sociological research technique for deniable acquisition of data that is often used to motivate DP~\cite{dwork2014algorithmic}.} Differential privacy is \emph{prima facie} useful as it provides mechanisms that limit information leakage from statistical database queries. However, the beneficial use of DP requires principled choice of $\epsilon$. Only then can a deployment provide a meaningful assurance of privacy to the individuals whose data are at risk.

An oft-used descriptive name for $\epsilon$ is the \emph{privacy budget}~\cite{dwork2006calibrating}, spent through information leakage in response to (a composed sequence of) queries. Like any budget, if too limited it has little utility as the returned statistics will not be meaningful. Conversely, if too generous nothing is beyond acquiring, and thus little to no data privacy is actually provided. It is a knob that must be tuned to trade off privacy and accuracy, with (i) no immediate from-first-principles approach with which to do so, as evidenced by the breadth of techniques proposed~\cite{abowd2019economic, garfinkel2018issues, dwork2010differential, dwork2014algorithmic, hsu2014differential, kohli2018epsilon, krehbiel2019choosing, lee2011much, laud2019interpreting, liu2019investigating, pejo2019together}, and (ii) penalties to accuracy or privacy from a flawed choice, as demonstrated by the controversy surrounding its use for disclosure avoidance in the 2020 US Census~\cite{abowd2019economic, cb, cb2, nyt, int, dpriv, dpriv2, garfinkel2018issues, kenny2021impact, ruggles2021role, santos2020differential, wood2017differential} and critical comments on deployments by Uber and Apple~\cite{mcsherry, tang2017privacy}.

\cut{There is a simple but helpful economic understanding of $\epsilon$, presented in~\cite{dwork2014algorithmic}. Suppose the running of a mechanism leads to some real-world outcome that an individual derives some profit or loss from, such as a job offer or a higher interest rate on a loan. The guarantee of differential privacy provides that the expected utility (the average profit or loss) for the individual cannot change by more than a factor of $e^{\epsilon}$ depending on whether or not their data are included in the analysis. When $\epsilon < 1$ then $e^{\epsilon} \approx 1 + \epsilon \approx 1$, and the utility can barely change. However, when $\epsilon$ is (much) bigger than one, suddenly this (admittedly very rough) upper bound on the worst-case change in utility can be immense. If, \emph{e.g.}, $\epsilon \approx 19.61$, then this $e^{\epsilon}$ factor is over three hundred million. Although this is just a bound, in practice it indicates that an individual may have no formal guarantee that their participation in the analysis will not drastically alter the likelihood of being harmed rather than helped by doing so. Invoking Tolstoy, Dwork \emph{et al.} in~\cite{dwork2019differential} formulate the maxim that `[w]hile all small $\epsilon$ are alike, each large $\epsilon$ is large after its own fashion, making it difficult to reason about them.' When large epsilons appear in practice, they demand scrutiny.}{}

\cut{The parameterization of a differentially private mechanism is not the only concern with its use, as the deployment of DP brings all the attendant difficulties of modeling. Any system that depends on statistical analysis or learning can -- whether through malice, negligence, or earnestness -- gild harms with a craven appeal to quantitative impartiality. Even assuming best intentions, data may be of insufficient quality or completeness for its proper use. This risk is made even more acute by the noisy nature of DP mechanisms, which require an excess of signal to survive. The use of DP for the 2020 US Census may, \emph{e.g.}, wipe small communities off the statistical map~\cite{nyt}. Which of the many variants and derivatives of DP is best suited for a given setting may also be a delicate decision. Moreover, by placing the locus of privacy on data analysis rather than collection or collation, even a socially beneficial use of differential privacy opens the door to the eventual misuse of that data in the future~\cite{cryptoeprint:2015:1162}.}{}\removeforshortfinal{\footnote{\cut{There do exist alternatives to the naive \emph{curation} model of differential privacy where a trusted party handles data collection and analysis, most notably the \emph{local}~\cite{dwork2014algorithmic, kasiviswanathan2011can} and \emph{shuffle}~\cite{bittau2017prochlo, cheu2019distributed, erlingsson2019amplification} models. Although they do not require complete trust in a single party, they have reduced accuracy, require additional data, and/or carry increased computational cost.}{}}} \cut{Finally, DP is designed to limit the marginal harm to a single individual from the decision to allow analysis of the individual's data but makes no promises about harm to the individual from trends at the population level. The canonical example, presented in depth in~\cite{dwork2014algorithmic}, concerns differentially private analysis relating health data and healthcare costs in order to set insurance premiums. If an individual has a behavior or condition that correlates with higher costs across the dataset, DP promises that the marginal inclusion of the individual's data in the analysis will not greatly exacerbate any premium increase attributable to the discovery of this relationship. However, it may be that, if every individual with the behavior or condition refused to participate in the analysis, the correlation would go unnoticed and the individual's premiums will not increase at all. Differential privacy minimizes the risk to an individual from the \emph{additional} inclusion of his or her data but does not necessarily minimize the risk to that individual from inferences made about them based on population-wide trends, to which their participation contributes. So even under DP the decision by an individual to share data as part of a collective action can still harm them. This point received particular attention during the debate over the 2020 US Census~\cite{dpriv}.}


But in the end, the implementer must make a choice of $\epsilon$. That choice is, in the opinion of Dwork and Smith, `essentially a social question'~\cite{dwork2010differential}. The problem has been studied formally through the lenses of resource allocation, economic valuations, and individual preferences~\cite{abowd2019economic, hsu2014differential, kohli2018epsilon, krehbiel2019choosing, pejo2019together}. These approaches read promisingly but generally require a responsible party -- either an individual or a corporation or government acting on the individual's behalf -- to precisely quantify privacy valuations or preferences over data and events. Whether this requirement is tenable is at best contentious in the prevailing research, especially for individuals who may struggle to comprehend the basic principles of differential privacy or assess the economic value of their data~\cite{acquisti2005privacy, acquisti2016economics, xiong2020towards}. The example of Uber also shows that even technically adept corporations may struggle with scientific evaluation of DP, let alone with earning the trust not to misuse data from its subjects~\cite{mcsherry, ohm2009broken}. Alternative research has explored using statistical techniques~\cite{laud2019interpreting, lee2011much, liu2019investigating}, which are however contingent on knowledge of some or all of an attacker's goals, prior beliefs, and auxiliary information. These requirements reverse one of the most lauded attributes of \textsc{Definition}~\ref{def:dp}, that the privacy guarantee of DP holds absolutely no matter the adversary or the data distribution~\cite{dwork2014algorithmic}.

Perhaps more realistic assessments of how $\epsilon$ will be chosen come from Dwork and Roth~\cite{dwork2014algorithmic}, from the experience of the US Census Bureau~\cite{abowd2019economic, cb, cb2, garfinkel2018issues}, and from  Dwork \emph{et al.} in \emph{Differential Privacy in Practice: Expose Your Epsilons!}~\cite{dwork2019differential}. In the first, after raising both technical and procedural, \emph{i.e.}, sociotechnical, difficulties with mathematically prescriptive techniques, the authors float an Epsilon Registry to `encourage better practices through transparency.' Although such a registry might very well develop principled defaults and standards in time, the need for such a process implicitly lays bare our inability to confidently make a privacy enhancing choice of $\epsilon$ for any given use case on its own terms. As for the US Census Bureau, the words of its own researchers in~\cite{garfinkel2018issues}, detailing a use of differential privacy predating its application to the census itself, are quite illuminating.
\begin{quote}
  \small
The value was set by having the practitioner prepare a set of graphs that showed the trade-off between privacy loss ($\epsilon$) and accuracy. The group then picked a value of $\epsilon$ that allowed for sufficient accuracy, then tripled it, so that the the researchers would be able to make several additional releases with the same data set without having to return to [the Data Stewardship Executive Policy committee] to get additional privacy-loss budget. The value of $\epsilon$ that was given out was far higher than those envisioned by the creators of differential privacy.
\end{quote}
Finally, in \emph{Expose Your Epsilons!} the authors revisit the registry, now motivated by a case study on how $\epsilon$ has been set by researchers and engineers in various early deployments of the technology. Although the study includes examples of survey respondents who, \emph{e.g.}, set their $\epsilon$ based on detailed threat modeling, Dwork \emph{et al.}  report selection methods similar to or even less principled than that of the Census Bureau, noting even that `there were practitioners who admitted the choice of $\epsilon$ was completely arbitrary without much consideration.' The authors conclude that `[i]n spite of the widespread preference of utility over privacy, there [is] no general agreement on how to choose $\epsilon$.'

In sum, there is a significant literature on the theoretical properties of $\epsilon$ and its principled choice under ideal circumstances, \emph{i.e.}, when some $0 < \epsilon \leq 1$ maintains sufficient utility. However, the theoretical, empirical, and policy-oriented literature on practically choosing large $\epsilon$ is far more limited, making it potentially difficult for practitioners to make a principled selection wile navigating our six hazards. And in practice, we can see specific ways in which the selection of $\epsilon$ traverses each of our hazards.

\paragraph{The Six Hazards: \#1 \& \#2.} For the first, it cannot be expected that someone without a mature mathematical background can read~\textsc{Definitions}~\ref{def:ei} and~\ref{def:dp} and immediately understand the sociotechnical implications of $\epsilon$. In their \emph{Differential Privacy: A Primer for a Non-Technical Audience}, Wood~\emph{et al.}~spend significant time introducing $\epsilon$, explaining its characteristics and implications, and experimentally demonstrating its effect~\cite{wood2017differential}. However, this discussion focuses on a choice of $\epsilon \in (0, 1]$ lying within the stiller theoretical waters where its consequences can be neatly characterized and explained. The authors leave to a footnote important real-world examples -- including those of Apple and the US Census Bureau -- that use large values of $\epsilon$ that fall significantly above that range of values addressed in the article body. As such, in its thoughtful pursuit of a balance between depth and accessibility the main text does not in the end illuminate the real complexity of the mathematical theory or available empirical evidence. As for our second hazard, in supporting documentation for a court filing defending its use of DP, the US Census Bureau highlights that it is `\emph{mathematically grounded} [emphasis added] in a way that allows statisticians to fully understand the limits of what they can make available and what kind of privacy they can provably offer'~\cite{cresp}. Although this statement is factually true, it emphasizes the formal nature of DP as the essential element, while omitting that those limits are a question of social choice. Although that mathematical grounding of DP allows an understanding, it does not itself provide one. The debate around the census itself demonstrates how the social elements of that understanding can be highly controversial.

\paragraph{The Six Hazards: \#3 \& \#4.} As for the third hazard, data privacy is of immense importance in the information society~\cite{ohm2009broken, wood2017differential}. But, DP and its variants -- among our most expressive and powerful techniques for providing it -- are dependent on careful management of this enigmatic parameterization. Although \cut{the simple utility analysis given previously provides accessible guidance as to why small epsilons are generally acceptable}{a simple utility analysis can show that small epsilons are generally acceptable}, the opaque consequences of large epsilons makes reasoning about the sociotechnical effects of their use much more difficult~\cite{dwork2014algorithmic, dwork2019differential}. An epsilon registry may provide a path towards consensus around choice for deployment, but the proposal of the registry itself belies our inability to always set $\epsilon$ for a given domain on its own terms. For our fourth hazard, choice of $\epsilon$ is unfalsifiable and indeterminate. What constitutes a privacy violation over statistical data cannot always be well defined, let alone reduced to an $\epsilon$. A sense of privacy is often intangible, may be communal as well as individual, and may change with time. Some might also consider a given privacy violation to be outweighed by the public good and so not a failure in a sociotechnical sense. Although we may be able to show that a given choice of $\epsilon$ does not prevent a specific attack over a specific data distribution in a specific adversarial context, when we strip away those caveats to a broader understanding of privacy violations, our ability to falsify a choice of $\epsilon$ melts away.

\paragraph{The Six Hazards: \#5 \& \#6.} For the fifth hazard, as noted in~\S\ref{sec:six}, work on setting $\epsilon$ in a principled way has lagged behind efforts to extend the theory of differential privacy. A survey that found hundreds of variants of DP in the literature covers the selection of $\epsilon$ in a short paragraph~\cite{desfontaines2020sok}. For the sixth and final, the importance of data privacy has made it a central focus of legislation and regulation of technology~\cite{GDPR, ohm2009broken}. Any attempt to use or manage DP in this context requires careful consideration of $\epsilon$ on which the actual privacy enhancement depends. Otherwise, these regulatory efforts will be greatly limited in precision, if not in influence.

\paragraph{A Last Comment.} We stress that these hazards are only just that. Their adverse effects are not inevitable. In both theory and practice -- as has been quite convincingly argued by its advocates in the case of disclosure avoidance -- differential privacy can provide accurate mechanisms buttressed by concrete privacy guarantees in a provable balance all prior approaches to the problem lack. Its deployments for the Census and beyond have almost certainly led to meaningful privacy gains.

\section{Explaining HMAPs}\label{sec:pa}

Many of our references originate in an ongoing effort by technically knowledgeable authors to proactively illuminate the scientific origins of socially consequential aspects of computation, \emph{e.g.},~\cite{abelson2015keys, barocas2016big, bamberger2021verification, kroll2016accountable, malik2020hierarchy, ohm2009broken, wood2017differential, weitzner2008information}. However, we find that such work is often structured as a primer and tends to be limited in both the breadth and depth of its scientific presentation. This limitation is natural and reasonable. An accessible narrative is essential for a non-technical audience. However, such writing may gloss over important points so as not to plunge an easy technical account into a morass of detail. Mathematical objects as small and peculiar as HMAPs may not be given due consideration. Proactive education about the nature of HMAPs requires a companion format that goes beyond such primers to catalogue -- as thoroughly as is reasonable -- every detail of a technology that impacts sociotechnical scrutiny.

An encouraging invention in modern computer-science research is the Systematization of Knowledge, or SoK, which has arisen within the applied-cryptography and information-security communities. Each of~\cite{barbosa2019sok, bonneau2015sok, cortier2016sok, desfontaines2020sok, fuller2017sok, papernot2018sok, thomas2021sok, unger2015sok} is an example on a relevant topic. The goal of these documents is not only to survey the current state of research within the domain but to relate and organize various designs and definitions in a way that surfaces dependency, inclusion, and variation among them. In other words, to go beyond a simple survey and characterize how ideas have developed and where they lie in relation to each other. The best SoKs are structured to allow a reader with neither deep nor broad knowledge of the field to assess the current state of affairs and jump to the work most relevant to them. The 2020 \emph{SoK: Differential Privacies} by Desfontaines and Pej{\'o} is an excellent example of a SoK with this structural clarity~\cite{desfontaines2020sok}. It provides seven axes along which differential privacy definitions vary, and it charts the relationships among the many proposals along them to allow readers to understand the surface of the space explored by DP research.

A possible companion to primers for educating about the nature of HMAPs would be a Systematization of Knowledge for Practitioners, or SoKfP. Such a document would survey and organize (i) the dimensions and trends in modeling of a given social or physical problem; (ii) their relationships to various published algorithmic techniques; and (iii) the dependence upon HMAPs of those designs. Unlike a traditional SoK, the purpose of the document would not be to give a systematic overview of scientific knowledge. Rather, the goal would be to capture as thoroughly as possible the mechanisms by which scientists have proposed computation as part of solving a social or physical problem. An essential element of this narrative is the interpretation of these techniques under the most influential empirical, humanist, and social-scientific perspectives. Organized and written at an appropriate level of detail (preferably by interdisciplinary scholars or in collaboration with knowledgeable practitioners), such a document could allow readers to quickly but carefully determine a broad picture of the technical methods by which computer scientists have proposed solving a problem. This analysis would consider the repercussions of their possible deployment as evaluated through a broad sociotechnical lens, with a particular aim towards effective analysis and governance.

Of course, to draft an effective SoKfP, authors must handle the challenge of first predicting the most relevant content for the document and then providing an intelligible, digestible, and ultimately useful systematization.  This is a highly nontrivial challenge, because the practitioners who are the intended readers are unlikely to have equivalent technical background to the authors. It may be easy for a SoKfP to `miss the mark' in either content or tone, addressing what researches believe practitioners should or would care about, rather than what they actually do. More generally, responsibility for the exchange of ideas and knowledge necessary for the effective governance of technology can never be a one-way street; practitioners would probably need to be active participants in the intellectual and professional environments that produce SoKfPs for the SoKfPs to succeed as communication tools. The development of interdisciplinary and research-industry-government venues such as the Fairness, Accountability, and Transparency (FAccT) and the Computer Science \& Law (CS\&LAW) conferences and communities -- as well as various supporting workshops and grants such as the NSF Designing Accountable Software Systems (DASS) program -- may very well provide ideal forums in which to produce the culture and collaborations needed for SoKfPs and other projects with related end goals.

\cut{Consider then, an ideal companion SoKfP to the work of Desfontaines and Pej{\'o} on differential privacy. It might begin by accounting for how compatible the underlying principle of modeling data privacy through indistinguishably-based data analysis is with perspectives from law, policy, economics, the social sciences, and the humanities, as in~\cite{ohm2009broken}. Any notable variants of DP -- especially those that respond to concerns of modeling or HMAPs with the standard definition -- could then be recounted with discussion as to when one such variant may be preferred. Careful description of these definitions in the mode of a primer similar to~\cite{wood2017differential} would then motivate $\epsilon$ and our understanding of its principled choice. This discussion would focus on -- and carefully distinguish between -- our theoretical, empirical, and social understandings of DP and $\epsilon$. Such a document would be valuable to an engineer looking for guidance on how to choose $\epsilon$ in deployment, a policymaker attempting to integrate DP into regulatory infrastructure, a lawyer trying to demonstrate harm from a loss of privacy, or a social scientist trying to interpret data analyzed under it.}{}

\emph{SoK: Hate, Harassment, and the Changing Landscape of Online Abuse} by Thomas \emph{et al.} is one of the closest works we are aware of in form and purpose to our conception of an SoKfP~\cite{thomas2021sok}. It combines research from various domains and methodologies, including a significant survey, to construct a taxonomy of hate and harassment in online spaces. However, as its focus lies less on the function of internet platform technologies and more on the social behaviors they enable, HMAPs mostly falls outside its scope. In a quite different way \emph{A Hierarchy of Limitations in Machine Learning} by Malik~\cite{malik2020hierarchy} furnishes another example of work similar in purpose to our proposed SoKfPs. It is mostly composed of qualitative methods, although written by a scientific expert who uses advanced mathematical tooling at points in the narrative. Malik is intentionally opinionated `making [it] a work ultimately of practice and of a practitioner rather than of analysis.' Additionally, it covers an immense scope that naturally limits the technical detail provided. Nonetheless, a synthesis that combined it with a technical SoK would exemplify the structure we view as potentially conducive to effective and broadly scoped presentation of urgent topics such as differential privacy, robust machine learning, or exceptional law-enforcement access.

Despite efforts by technical researchers to proactively produce scholarship helpful to practitioners, there will also always be a reactive role for computer scientists in responding to scrutiny from without, such as from journalists or during expert testimony. Every query or line of inquiry practitioners may raise cannot be predicted ahead of time. Even if scientists could, systematizing it all would be an impractical undertaking. How best to convey the nuance of HMAPs in an \emph{ad hoc} manner will likely be specific to the exact combination of computer-science discipline and the expertise of the questioner. Nonetheless, our examples and hazards highlight generic attributes of HMAPs whose regularized treatment may broadly help sociotechnical analysis. A SoKfP provides the opportunity to organize and present our knowledge in a manner uniquely tailored for use by specific communities of practitioners. These generic attributes are far less structured but are at least intuitively understandable.

Most importantly, we consider the following loose classification of HMAPs. It is intended to give a simple context to how an object originates and the extent to which computer scientists understand its acceptance or selection. A heuristic model, assumption, or parameter may be:
\begin{enumerate}[(i)]
    \item \emph{formal}: the object has a well defined mathematical truth that has not been conclusively established or is being used in spite of its known falsity; \emph{e.g.}, heuristic models and cryptographic hardness assumptions~\cite{goldwasser2003security, goldreich2009foundations, naor2003cryptographic};
    \item \emph{formal--empirical}: a formal mathematical argument provides a basis on which to conduct empirical analysis; \emph{e.g.}, cryptographic key sizes (security parameters)~\cite{alagic2022status, chang2012third, katz2014introduction}, the effect of $\epsilon$ on the accuracy of a differential privacy mechanism~\cite{dwork2006calibrating, dwork2008differential, dwork2009differential, dwork2014algorithmic, wood2017differential}, sample complexity~\cite{shalev2014understanding};
    \item \emph{physical}: the object represents a physical process that can be evaluated empirically by engineers or natural scientists; \emph{e.g.}, assumptions about sensors in cyberphysical systems~\cite{ntsb}; choice of $\epsilon$ in robust ML when the modeled process is physical~\cite{goodfellow2014explaining, madry2017towards, tsipras2018robustness};
    \item \emph{social--empirical}: the validity or choice of object is a social question, but it is one that can be placed in empirical terms by social scientists; \emph{e.g.}, choice of $\epsilon$ in robust ML when the modeled process is social~\cite{goodfellow2014explaining, madry2017towards, tsipras2018robustness}, choice of $\epsilon$ in differential privacy following the work quantifying privacy in economic terms~\cite{abowd2019economic, hsu2014differential, kohli2018epsilon, krehbiel2019choosing, pejo2019together}; choice of fairness parameters under empirical advisement from social-scientific research~\cite{celis2017ranking, dwork2012fairness, selbst2019fairness}; and
    \item \emph{social}: the validity or choice of object is a social question that can only be resolved by humanist analysis of its effects; \emph{e.g.}, all social--empirical examples when we do not have satisfactory social-scientific methods.
\end{enumerate}
The distinction between the last two cannot be resolved by computer scientists alone, but how naturally social-scientific concepts map onto computational counterparts is an essential question that technical researchers must help analyze. We note as well a few other dynamics within our classification. Since real physical objects can only be evaluated empirically, we do not need to distinguish between \emph{physical} and \emph{physical--empirical} HMAPs. Of course, physical objects can be formally modeled and that formal model may also be evaluated, in which case any HMAPs relevant to that model would be \emph{formal} or \emph{formal--empirical} as appropriate. For example, the correctness of physical sensor used on the Boeing 737-MAX aircraft in operation is a physical assumption, but for an attempt to formally verify the software system that incorporates the readings of the sensor its correctness would be a formal assumption instead. Separately, we note that a formal model can be of a realized (or just realizable) physical system, social system, or of both or of neither. It is the method of reasoning that is relevant to the classification, not the system being reasoned about. So the HMAPs used by a formal model of a social system would be classified as \emph{formal} since the mode of reasoning in use is axiomatic, rather than, \textit{e.g.}, statistical (as in the case of \emph{social--empirical}) or humanistic (in the case of \emph{social}).

To this above classification we can also add four additional markers. First, whether the object is indeterminate. Second, whether it is unfalsifiable. Third, whether it receives active research attention and has a robust scientific literature. And fourth, whether there exists engineering experience with its development and deployment. For many practitioners, characterizations of HMAPs -- even without the standardized structure of a SoKfP -- along these high-level lines may be more helpful than attempts to demonstrate their function through mathematical descriptions, toy examples, or results from irreconcilable domains.

\cut{\section{Conclusion}\label{sec:conc}

We have discussed how heuristic models, assumptions, and parameters -- the HMAPs -- contribute to the sociotechnical dimensions of computation. Our thesis is, in essence, that the small and secondary nature of these objects does not mitigate their hazard to our understanding of how programs and protocols interact with society and the world. Our proposals in~\S\ref{sec:pa} may, we hope, stimulate computer scientists to consider how best to address sociotechnical concerns with HMAPs in a manner accessible to practitioners. Regardless, it is essential that computer scientists address how to consistently provide the full measure of designs to those engaged in their principled use within society.}{}